\documentclass[aps,preprint,nofootinbib]{revtex4}%
\usepackage{amsfonts}
\usepackage{amsmath}
\usepackage{amssymb}
\usepackage{graphicx}%
\providecommand{\U}[1]{\protect\rule{.1in}{.1in}}

\begin{document}
\preprint{ }
\title[Short title for running header]{On the gauge symmetries of the spinning particle}
\author{N. Kiriushcheva}
\email{nkiriush@uwo.ca}
\affiliation{The Department of Applied Mathematics, The University of Western Ontario,
London, Ontario, N6A 5B7, Canada}
\author{S.V. Kuzmin}
\email{skuzmin@uwo.ca}
\affiliation{The Department of Economics, Business, and Mathematics, The King's University
College, London, Ontario, N6A 2M3, Canada}
\author{D.G.C. McKeon}
\email{dgmckeo2@uwo.ca}
\affiliation{The Department of Applied Mathematics, The University of Western Ontario,
London, Ontario, N6A 5B7, Canada}
\affiliation{The Department of Mathematics and Computer Science, Algoma University, Sault
St. Marie, Ontario, P6A 2G4, Canada}

\begin{abstract}
We reconsider the gauge symmetries of the spinning particle by a direct
examination of the Lagrangian using a systematic procedure based on the
Noether identities. It proves possible to find a set of local Bosonic and
Fermionic gauge transformations that have a simple gauge group structure,
which is a true Lie algebra, both for the massless and massive case. This new
Fermionic gauge transformation of the \textquotedblleft
position\textquotedblright\ and \textquotedblleft spin\textquotedblright%
\ variables in the action decouples from that of the \textquotedblleft
einbein\textquotedblright\ and \textquotedblleft gravitino\textquotedblright.

It is also possible to redefine the fields so that this simple algebra of
commutators of the gauge transformations can be derived directly starting from
the Lagrangian written in these new variables.

We discuss a possible extension of our analysis of this simple model to more
complicated cases.

\end{abstract}
\volumeyear{year}
\volumenumber{number}
\issuenumber{number}
\eid{identifier}
\maketitle

\section{Introduction}

For four decades, supersymmetry has been studied intensively. The local
version of this symmetry, supergravity, is most easily realized by the
spinning particle \cite{BrinkPLB, BrinkNPB, BrinkPLB-err}; this is
supergravity theory in 0+1 dimensions.

In the original presentation of the action for the spinning particle, a
particular set of local Bosonic and Fermionic gauge transformations was given
\cite{BrinkPLB, BrinkNPB}; their form appears to be motivated by the
supersymmetric and diffeomorphism gauge transformations present in the
supergravity action in 3+1 dimensions \cite{Niew}. However, as was noted in
\cite{BrinkNPB}, these gauge transformations do not have a gauge group
structure in which the structure functions are field independent.

We wish in this paper to point out that this deficiency can be overcome by
altering the form of the gauge transformations in a simple way. This is
systematically done by direct derivation starting from the Lagrangian. A
general form of an arbitrary gauge transformation can be derived from
differential identities (DIs), which are linear combinations of Euler-Lagrange
derivatives (ELDs) of the action; this method can be applied to any action
with a known gauge transformation to search for a form of the local gauge
transformations that simplifies the gauge group properties.

This general expression for a gauge transformation obtained from a DI of the
action for the spinning particle can also be used to find a reparametrization
of the fields so that the Fermionic gauge transformation decouples the
\textquotedblleft position\textquotedblright\ and \textquotedblleft
spin\textquotedblright\ fields from the \textquotedblleft
einbein\textquotedblright\ and \textquotedblleft gravitino\textquotedblright\ fields.

We note that the gauge symmetry structure of the spinning particle action can
also be studied using the canonical structure of the action; a generator of
both Bosonic and Fermionic gauge transformations that have a simple gauge
group structure can be derived from first class constraints \cite{Gerry}. The
same procedure can be applied for the superparticle action \cite{Gerry-arxiv}.

\section{The spinning particle}

We start by examining the general case of a particle action%

\begin{equation}
S=\int d\tau L\left(  q_{i}\left(  \tau\right)  ,\dot{q}_{i}\left(
\tau\right)  \right)  \label{a1}%
\end{equation}

(where $\dot{q}_{i}\left(  \tau\right)  \equiv\frac{\partial q_{i}}%
{\partial\tau}$) and considering a variation of each of the fields
$q_{i}\left(  \tau\right)  $ so that%

\begin{equation}
\delta S=\int d\tau\sum_{i}\delta q_{i}\left(  \tau\right)  E_{q_{i}}.
\label{2}%
\end{equation}

In eq. (\ref{2}), the Euler-Lagrange derivatives (ELDs) $E_{q_{i}}$ are given by%

\begin{equation}
E_{q_{i}}=\frac{\delta L}{\delta q_{i}}-\partial\frac{\delta L}{\delta\dot
{q}_{i}} \label{3}%
\end{equation}

(where $\partial\equiv\frac{\partial}{\partial\tau}$). If this were to vanish
for arbitrary $\delta q_{i}$, then we have the fields $q_{i}$ satisfying the
equation of motion $E_{q_{i}}=0$. However, if the form of $\delta q_{i}$ is
such that $\delta S=0$ for arbitrary $q_{i}\left(  \tau\right)  $, then we
have a local gauge symmetry of the action $S$. According to the Noether
theorem \cite{Noether, Noether-eng}, any gauge symmetry $q_{i}\rightarrow
q_{i}+\delta q_{i}$ of $S$ satisfies the equation%

\begin{equation}
\sum_{i}\delta q_{i}E_{q_{i}}\equiv0, \label{4}%
\end{equation}

which leads to the differential identity $I$ from%

\begin{equation}
I\alpha=\sum_{i}\delta q_{i}\left(  \tau\right)  E_{q_{i}}, \label{E5}%
\end{equation}

where $\alpha$ is a gauge parameter.

The Lagrangian we are particularly concerned with is that of the spinning particle%

\begin{equation}
L=\frac{1}{2}\left[  e^{-1}\dot{\phi}^{\mu}\dot{\phi}_{\mu}-i\psi^{\mu}%
\dot{\psi}_{\mu}-ie^{-1}\chi\psi^{\mu}\dot{\phi}_{\mu}\right]  , \label{L1}%
\end{equation}

where $\phi^{\mu}\left(  \tau\right)  $ and $e\left(  \tau\right)  $ are two
Bosonic fields and $\psi^{\mu}\left(  \tau\right)  $ and $\chi\left(
\tau\right)  $ are two Fermionic fields, or two pairs of superpartners
$\left(  e,\chi\right)  $ and$\left(  \phi_{\mu},\psi^{\mu}\right)  $. The
index $\mu$ in $\phi^{\mu}$ and $\psi^{\mu}$ is raised and lowered by the
Minkowski metric tensor.

In refs. \cite{BrinkPLB, BrinkNPB}, the Bosonic gauge invariance present in
this action is given by the \textquotedblleft diffeomorphism\textquotedblright\ transformation%

\begin{equation}
\delta_{f}e=\partial\left(  fe\right)  ,\qquad\delta_{f}\chi=\partial\left(
f\chi\right)  ,\qquad\delta_{f}\phi^{\mu}=f\partial\left(  \phi^{\mu}\right)
,\qquad\delta_{f}\psi^{\mu}=f\partial\left(  \psi^{\mu}\right)  ,
\label{f-tr1}%
\end{equation}

while the Fermionic or supersymmetry transformation was chosen by the authors
of \cite{BrinkPLB, BrinkNPB} as%

\begin{equation}
\delta_{\alpha}\phi^{\mu}=i\alpha\psi^{\mu},\qquad\delta_{\alpha}e=i\alpha
\chi,\qquad\delta_{\alpha}\chi=2\dot{\alpha},\qquad\delta_{\alpha}\psi^{\mu
}=\alpha e^{-1}\left(  \dot{\phi}^{\mu}-\frac{i}{2}\chi\psi^{\mu}\right)
\label{super-tr1}%
\end{equation}

with $f\left(  \tau\right)  $ and $\alpha\left(  \tau\right)  $ being Bosonic
and Fermionic gauge parameters, respectively.

Under the transformations of (\ref{f-tr1}) and (\ref{super-tr1}) we find%

\begin{equation}
\delta_{f}L=\partial\left(  fL\right)  , \label{L1tr}%
\end{equation}

\begin{equation}
\delta_{\alpha}L=\partial\left(  \frac{i}{2e}\alpha\psi^{\mu}\dot{\phi}_{\mu
}\right)  , \label{L1super-tr}%
\end{equation}

respectively.

The group property of the Bosonic transformation of (\ref{f-tr1}) can be
combined into one equation%

\begin{equation}
\left[  \delta_{f_{1}},\delta_{f_{2}}\right]  \left(  \phi^{\mu},\psi^{\mu
},e,\chi\right)  =\delta_{\tilde{f}}\left(  \phi^{\mu},\psi^{\mu}%
,e,\chi\right)  , \label{eqnS10}%
\end{equation}

where%

\begin{equation}
\tilde{f}=f_{2}\dot{f}_{1}-f_{1}\dot{f}_{2} \label{eqnS11}%
\end{equation}

and so the structure functions of this transformation are field independent.
The same is valid also for the commutator of Bosonic and Fermionic transformations%

\begin{equation}
\left[  \delta_{f},\delta_{\alpha}\right]  \left(  \phi^{\mu},\psi^{\mu
},e,\chi\right)  =\delta_{\tilde{\alpha}}\left(  \phi^{\mu},\psi^{\mu}%
,e,\chi\right)  \label{e7}%
\end{equation}

with the parameter%

\begin{equation}
\tilde{\alpha}=-f\dot{\alpha}. \label{e8}%
\end{equation}

However, when commuting two Fermionic gauge transformations we find from eqs.
(\ref{f-tr1}) and (\ref{super-tr1})%

\begin{equation}
\left[  \delta_{\alpha_{1}},\delta_{\alpha_{2}}\right]  \left(  \phi^{\mu
},\psi^{\mu},e,\chi\right)  =\delta_{\bar{f}}\left(  \phi^{\mu},\psi^{\mu
},e,\chi\right)  +\delta_{\bar{\alpha}}\left(  \phi^{\mu},\psi^{\mu}%
,e,\chi\right)  , \label{e5}%
\end{equation}

where%

\begin{equation}
\bar{f}=-\frac{2i\alpha_{1}\alpha_{2}}{e}\text{ \ \ and \ \ \ }\bar{\alpha
}=-\frac{1}{2}\bar{f}\chi. \label{e6}%
\end{equation}

We thus see that the transformations of eqs. (\ref{f-tr1}) and
(\ref{super-tr1}) do not possess a simple group property, as was noticed by
the authors of \cite{BrinkNPB}, because of the explicit dependence of the
gauge parameters in eq. (\ref{e6}) on the fields.

We will now exploit eq. (\ref{4}), using the Noether identities, to find a
form of gauge transformations that has a simpler gauge group. This method was
outlined in \cite{transl, field-param}. We begin from the ELDs of the
Lagrangian (\ref{L1})%

\begin{equation}
E_{e}=\frac{\delta L}{\delta e}=-\frac{1}{2e^{2}}\left(  \dot{\phi}^{\mu}%
\dot{\phi}_{\mu}-i\chi\psi^{\mu}\dot{\phi}_{\mu}\right)  , \label{E1}%
\end{equation}

\begin{equation}
E_{\chi}=\frac{\delta L}{\delta\chi}=-\frac{i}{2e}\psi^{\mu}\dot{\phi}_{\mu},
\label{E2}%
\end{equation}

\begin{equation}
E_{\phi\mu}=\frac{\delta L}{\delta\phi^{\mu}}=-\partial\left(  e^{-1}\dot
{\phi}_{\mu}-\frac{i}{2e}\chi\psi_{\mu}\right)  , \label{E3}%
\end{equation}

\begin{equation}
E_{\psi\mu}=\frac{\delta L}{\delta\psi^{\mu}}=-i\dot{\psi}_{\mu}+\frac{i}%
{2e}\chi\dot{\phi}_{\mu}, \label{E4}%
\end{equation}

and so by eqs. (\ref{f-tr1}) and (\ref{super-tr1}) respectively we have the
Bosonic DI%

\begin{equation}
I=-e\partial E_{e}-\chi\partial E_{\chi}+\partial\left(  \phi^{\mu}\right)
E_{\phi\mu}+\partial\left(  \psi^{\mu}\right)  E_{\psi\mu}\equiv0
\label{eqnS15}%
\end{equation}

and the Fermionic DI%

\begin{equation}
\Psi=-2\partial E_{\chi}+i\chi E_{e}+i\psi^{\mu}E_{\phi\mu}+\left(  \dot{\phi
}^{\mu}-\frac{i}{2}\chi\psi^{\mu}\right)  e^{-1}E_{\psi\mu}\equiv0.
\label{eqnS17}%
\end{equation}

We now use the fact that any linear combination of these two DIs yields a DI;
the number of linearly independent DIs cannot be changed \cite{Noether,
Noether-eng}. The simplest modification is to multiply eq. (\ref{eqnS17}) by a
function $h\left(  e\right)  $, if we want to preserve its tensorial and
Fermionic nature; the resulting DI corresponds to the gauge transformations%

\begin{equation}
\delta_{\alpha}\chi=\partial\left(  2\alpha h\left(  e\right)  \right)
,\qquad\delta_{\alpha}e=i\alpha\chi h\left(  e\right)  ,\quad\label{eqnS19}%
\end{equation}

\begin{equation}
\delta_{\alpha}\phi^{\mu}=i\alpha\psi^{\mu}h\left(  e\right)  ,\qquad
\delta_{\alpha}\psi^{\mu}=\alpha e^{-1}\left(  \dot{\phi}^{\mu}-\frac{i}%
{2}\chi\psi^{\mu}\right)  h\left(  e\right)  , \label{eqnS20}%
\end{equation}

which gives the commutator%

\begin{equation}
\left[  \delta_{\alpha_{1}},\delta_{\alpha_{2}}\right]  \chi=\partial\left(
-4i\alpha_{1}\alpha_{2}\frac{dh\left(  e\right)  }{de}h\left(  e\right)
\chi\right)  . \label{eqnS22}%
\end{equation}

All field dependence in this commutator disappears if we take%

\begin{equation}
h\left(  e\right)  =\sqrt{e}, \label{eqnS24}%
\end{equation}

so that the Fermionic gauge transformation becomes%

\begin{equation}
\delta_{\alpha}\phi^{\mu}=i\sqrt{e}\alpha\psi^{\mu},\qquad\delta_{\alpha
}e=i\sqrt{e}\alpha\chi,\qquad\delta_{\alpha}\chi=2\partial\left(  \sqrt
{e}\alpha\right)  ,\qquad\delta_{\alpha}\psi^{\mu}=\alpha\frac{1}{\sqrt{e}%
}\left(  \dot{\phi}^{\mu}-\frac{i}{2}\chi\psi^{\mu}\right)  . \label{eqnS26}%
\end{equation}

The commutator of two new Fermionic gauge transformations is now becomes%

\begin{equation}
\left[  \delta_{\alpha_{1}},\delta_{\alpha_{2}}\right]  \left(  \phi^{\mu
},\psi^{\mu},e,\chi\right)  =\delta_{\bar{f}}\left(  \phi^{\mu},\psi^{\mu
},e,\chi\right)  \label{eqnS30}%
\end{equation}

with%

\begin{equation}
\bar{f}=-2i\alpha_{1}\alpha_{2}, \label{eqnS31}%
\end{equation}

while together the Fermionic and Bosonic gauge transformations result in%

\begin{equation}
\left[  \delta_{f},\delta_{\alpha}\right]  \left(  \phi^{\mu},\psi^{\mu
},e,\chi\right)  =\delta_{\tilde{\alpha}}\left(  \phi^{\mu},\psi^{\mu}%
,e,\chi\right)  \label{eqnS40}%
\end{equation}

with%

\begin{equation}
\tilde{\alpha}=f\dot{\alpha}-\frac{1}{2}\alpha\dot{f}. \label{eqnS41}%
\end{equation}

With the gauge transformations of eqs. (\ref{eqnS10}) and (\ref{eqnS26}) we
see that we have a gauge algebra whose structure functions are independent of
fields; this simple algebra automatically satisfies the Jacobi identities.

One can supplement the Lagrangian of eq. (\ref{L1}) with a \textquotedblleft
mass\textquotedblright, or \textquotedblleft cosmological\textquotedblright\ term%

\begin{equation}
L_{5}=\frac{1}{2}\left(  m^{2}e+i\psi_{5}\dot{\psi}_{5}-im\psi_{5}\chi\right)
. \label{eqnS60}%
\end{equation}

The Lagrangian $L_{5}$ of eq. (\ref{eqnS60}) is invariant under the gauge
transformations of eqs. (\ref{f-tr1}) and (\ref{super-tr1}) provided we also
have \cite{BrinkPLB}%

\begin{equation}
\delta_{f}\psi_{5}=f\dot{\psi}_{5} \label{eqnS61a}%
\end{equation}

and%

\begin{equation}
\delta_{\alpha}\psi_{5}=m\alpha, \label{eqnS62a}%
\end{equation}

which result in%

\begin{equation}
\delta_{f}L_{5}=\partial\left(  fL_{5}\right)  ,\text{\qquad}\delta_{\alpha
}L_{5}=\partial\left(  \alpha\frac{i}{2}m\psi_{5}\right)  ; \label{eqnS62}%
\end{equation}

and the commutator of two Fermionic transformations is very simple:%

\begin{equation}
\left[  \delta_{\alpha_{1}},\delta_{\alpha_{2}}\right]  \psi_{5}=0.
\label{eqnS62b}%
\end{equation}

However, the transformation of eq. (\ref{eqnS62a}) has been supplemented by an
extra piece given in \cite{BrinkNPB, BrinkPLB-err} so that%

\begin{equation}
\delta_{\alpha}\psi_{5}=m\alpha+\frac{i}{me}\alpha\psi_{5}\left(  \dot{\psi
}_{5}-\frac{1}{2}m\chi\right)  \label{eqnS61}%
\end{equation}

in order that the gauge algebra of eq. (\ref{e5}) is satisfied.

The ELD associated with $\psi_{5}$ is%

\begin{equation}
E_{\psi_{5}}=i\dot{\psi}_{5}-\frac{i}{2}m\chi; \label{eqnS65}%
\end{equation}

when this is combined with the Fermionic gauge transformation of eq.
(\ref{eqnS61}) we end up with the DI%

\begin{equation}
\Psi\Rightarrow\Psi+\left[  m+\frac{i}{me}\psi_{5}\left(  \dot{\psi}_{5}%
-\frac{1}{2}m\chi\right)  \right]  E_{\psi_{5}}\equiv0, \label{eqnS67}%
\end{equation}

where $\Psi$ is given by eq. (\ref{eqnS17}) with new \textquotedblleft
einbein\textquotedblright\ and \textquotedblleft gravitino\textquotedblright%
\ ELDs are%

\begin{equation}
E_{e}\Rightarrow E_{e}+\frac{1}{2}m^{2}, \label{eqnS63}%
\end{equation}

\begin{equation}
E_{\chi}\Rightarrow E_{\chi}+\frac{i}{2}m\psi_{5}.\label{eqnS64}%
\end{equation}

We would like to note that, despite of consistency with the gauge algebra of
eq. (\ref{e5}), the gauge transformation (\ref{eqnS61}) with the extra piece
is not legitimate because it introduces a term in the DI proportional to the
square of ELD $E_{\psi_{5}}$, as eq. (\ref{eqnS67}) can be written in the form%

\begin{equation}
\Psi+mE_{\psi_{5}}+\frac{1}{me}\psi_{5}\left(  E_{\psi_{5}}\right)  ^{2}%
\equiv0.\label{eqnS67a}%
\end{equation}

It contradicts the definition of a DI as being a linear combination of ELDs.

We now make use of the function $h\left(  e\right)  $ of eq. (\ref{eqnS24})
which modifies the DI so that we have the Fermionic gauge transformation%

\begin{equation}
\delta_{\alpha}\psi_{5}=\sqrt{e}m\alpha+\frac{i}{m\sqrt{e}}\alpha\psi
_{5}\left(  \dot{\psi}_{5}-\frac{1}{2}m\chi\right)  . \label{eqnS68}%
\end{equation}

This is consistent with the gauge algebra given by eqs. (\ref{eqnS30}%
-\ref{eqnS41}).

One can also make an invertible change of variables in the original action
without destroying the gauge algebra. For example, in ref. \cite{BrinkPLB} a
rescaling of Fermionic variables%

\begin{equation}
\psi^{\mu}=\frac{1}{\sqrt{e}}\tilde{\psi}^{\mu},\qquad\chi=\frac{1}{\sqrt{e}%
}\tilde{\chi}, \label{change1}%
\end{equation}

\begin{equation}
\psi_{5}=\sqrt{e}\tilde{\psi}_{5} \label{change1-5}%
\end{equation}

leads to the actions of eqs. (\ref{L1}, \ref{eqnS60}) being replaced by%

\begin{equation}
\tilde{L}=\frac{1}{2}\left[  e^{-1}\dot{\phi}^{\mu}\dot{\phi}_{\mu}%
-ie^{-1}\tilde{\psi}^{\mu}\partial\left(  \tilde{\psi}_{\mu}\right)
-ie^{-2}\tilde{\chi}\tilde{\psi}^{\mu}\dot{\phi}_{\mu}\right]  \label{L2}%
\end{equation}

and%

\begin{equation}
\tilde{L}_{5}=\frac{1}{2}\left(  m^{2}e+ie\tilde{\psi}_{5}\partial\left(
\tilde{\psi}_{5}\right)  -im\tilde{\psi}_{5}\tilde{\chi}\right)  .
\label{L2-5}%
\end{equation}

One can easily obtain the DI in terms of these new variables. The resulting
gauge transformations for these new variables are%

\begin{equation}
\delta_{f}\tilde{\chi}=\partial\left(  f\tilde{\chi}\right)  +\frac{1}{2}%
\dot{f}\tilde{\chi},\qquad\delta_{f}\tilde{\psi}^{\mu}=f\partial\left(
\tilde{\psi}^{\mu}\right)  +\frac{1}{2}\dot{f}\tilde{\psi}^{\mu},\qquad
\delta_{f}\tilde{\psi}_{5}=f\partial\left(  \tilde{\psi}_{5}\right)  -\frac
{1}{2}\dot{f}\tilde{\psi}_{5} \label{f-tr2}%
\end{equation}

for the Bosonic case (the gauge transformations $\delta_{f}e$ and $\delta
_{f}\phi^{\mu}$ remains the same as in (\ref{f-tr1})), and%

\begin{equation}
\delta_{\alpha}e=i\alpha\tilde{\chi},\qquad\delta_{\alpha}\tilde{\chi}%
=2\dot{\alpha}e+\alpha\dot{e}, \label{super-tr2}%
\end{equation}

\begin{equation}
\delta_{\alpha}\phi^{\mu}=i\alpha\tilde{\psi}^{\mu},\qquad\delta_{\alpha
}\tilde{\psi}^{\mu}=\alpha\dot{\phi}^{\mu}, \label{super-tr3}%
\end{equation}

\begin{equation}
\delta_{\alpha}\tilde{\psi}_{5}=m\alpha+\frac{i}{m}\alpha\tilde{\psi}%
_{5}\partial\left(  \tilde{\psi}_{5}\right)  \qquad\label{super-tr5}%
\end{equation}

for the Fermionic case. The Fermionic transformations of eq. (\ref{super-tr2},
\ref{super-tr3}, \ref{super-tr5}) are much simpler than those of
(\ref{super-tr1}, \ref{eqnS68}). In addition, the transformations of
$\phi^{\mu}$ and $\tilde{\psi}^{\mu}$ (the \textquotedblleft
position\textquotedblright\ and \textquotedblleft spin\textquotedblright%
\ fields) decouple from those of $e$ and $\tilde{\chi}$ (the \textquotedblleft
einbein\textquotedblright\ and \textquotedblleft gravitino\textquotedblright%
\ fields), as well as $\tilde{\psi}_{5}$ transforms separately from other fields.

Despite this new form of the gauge transformations, we retain the simple gauge
algebra of eqs. (\ref{eqnS10}, \ref{eqnS11}, \ref{eqnS30}-\ref{eqnS41}) in
which all structure functions are field independent:%

\begin{equation}
\left[  \delta_{f_{1}},\delta_{f_{2}}\right]  \text{ }field=\delta_{\tilde{f}%
}\text{ }field\text{,\qquad with \ \ \ }\tilde{f}=f_{2}\dot{f}_{1}-f_{1}%
\dot{f}_{2}, \label{c1}%
\end{equation}

\begin{equation}
\left[  \delta_{\alpha_{1}},\delta_{\alpha_{2}}\right]  \text{ }%
field=\delta_{\bar{f}}\text{ }field\text{,\qquad with \ \ \ \ }\bar
{f}=-2i\alpha_{1}\alpha_{2}, \label{c2}%
\end{equation}

\begin{equation}
\left[  \delta_{f},\delta_{\alpha}\right]  \text{ }field=\delta_{\tilde
{\alpha}}\text{ }field\text{,\qquad with \ \ \ \ }\tilde{\alpha}=f\dot{\alpha
}-\frac{1}{2}\alpha\dot{f}. \label{c3}%
\end{equation}

The Jacobi identities automatically hold for such gauge transformations.

A particularly simple form of the gauge transformations for the fields which
almost trivializes the calculations of the gauge algebra can be obtained by
replacing $e$ by $g$ and $\tilde{\chi}$ by $\chi^{\prime\prime}$ where%

\begin{equation}
e=\exp\left(  g\right)  , \label{p-4}%
\end{equation}

\begin{equation}
\chi^{\prime\prime}=\exp\left(  -g\right)  \tilde{\chi}. \label{p-8}%
\end{equation}

We now find the Bosonic transformation%

\begin{equation}
\delta_{f}g=\dot{f}+f\dot{g}, \label{p-5}%
\end{equation}

\begin{equation}
\delta_{f}\chi^{\prime\prime}=f\dot{\chi}^{\prime\prime}+\frac{1}{2}\dot
{f}\chi^{\prime\prime} \label{p-11}%
\end{equation}

and the Fermionic one%

\begin{equation}
\delta_{\alpha}\tilde{\psi}^{\mu}=\alpha\dot{\phi}^{\mu},\qquad\delta_{\alpha
}\phi^{\mu}=i\alpha\tilde{\psi}^{\mu}, \label{p-9}%
\end{equation}

\begin{equation}
\delta_{\alpha}g=i\alpha\chi^{\prime\prime},\qquad\delta_{\alpha}\chi
^{\prime\prime}=2\dot{\alpha}+\alpha\dot{g}. \label{p-10}%
\end{equation}

This last parametrization (\ref{p-4}, \ref{p-8}) has especially simple
transformations that makes calculation of commutators of two supersymmetry
transformations almost trivial.

\section{Discussion}

\bigskip

By working directly with the DIs obtained from the action for the spinning
particle, we have derived a set of Bosonic (B) and Fermionic (F) gauge
transformations that have a simple gauge algebra of the form%

\begin{equation}
\left[  B,B\right]  =B,\qquad\left[  F,F\right]  =B,\qquad\left[  F,B\right]
=F.\mathit{\ } \label{p-12}%
\end{equation}

In this algebra, all structure functions are field independent and the Jacobi
identity is satisfied. This is an improvement over the original set of gauge
transformations appearing in refs. \cite{BrinkPLB, BrinkNPB}. Note, if we are
seeking for gauge transformations of Bosonic and Fermionic fields that form a
Lie algebra with field independent structure functions, then the only form
possible is that of eq. (\ref{p-12}).

The actual form of the gauge transformations has been simplified through a
field redefinition while retaining the simple algebra of (\ref{p-12}) for the
gauge transformations.

We would like to investigate more complicated models that have a local
Fermionic symmetry to see if similar simplifications can be effected. An
$O\left(  N\right)  $ generalization of the spinning particle, the spinning
string and supergravity in $D\geq3$ dimensions should all be examined with
this objective in mind.

A general problem would be to establish the relationship between the DI of eq.
(\ref{4}) that is satisfied by any gauge transformation and the gauge
generator obtained from the first class constraints arising from the canonical
structure of the theory \cite{Castellani, HTZ}.

The gauge generator derived from the first class constraints can be used to
determine a gauge invariance of a theory (even one previously unsuspected, as
in the case of the first order Einstein-Hilbert action in two dimensions
\cite{2D}). However, it has not as yet proven possible to determine directly
from the Lagrangian and its ELDs all independent DIs of the form (\ref{4}) and
consequently, all gauge symmetries, though once one has a gauge
transformation, alternate gauge transformations can easily be found by using
this DI.

\bigskip

\section{Acknowledgement}

Roger Macleod provided a helpful suggestion.


\bigskip

\begin{thebibliography}{99}                                                                                               %


\bibitem {BrinkPLB}L. Brink, S. Deser, B. Zumino, P. Di Vecchia, P. Howe,
Phys. Lett. B 64 (1976) 435-438

\bibitem {BrinkNPB}L. Brink, P. Di Vecchia, P. Howe, Nucl. Phys. B 118 (1977) 76-94

\bibitem {BrinkPLB-err}Erratum, Phys. Lett. B 68 (1977) 488

\bibitem {Niew}P. van Niewenhuizen, Phys. Rep. 68 (1981) 189

\bibitem {Gerry}D.G.C. McKeon, Can. J. Phys. 90 (2012) 701-705

\bibitem {Gerry-arxiv}D.G.C. McKeon, arXiv:1209.4909 [hep-th]

\bibitem {Noether}Noether, E.: Nachr. d. K\"{o}nig. Gesellsch. d. Wiss. zu
G\"{o}ttingen, Math.-phys. Klasse, 235 (1918)

\bibitem {Noether-eng}Noether, E. (M.A. Tavel's English translation): arXiv:physics/0503066

\bibitem {transl}N. Kiriushcheva and S. V. Kuzmin, Gen. Rel. Grav. 42 (2010) 2613-2631

\bibitem {field-param}N. Kiriushcheva, P. G. Komorowski, and S. V. Kuzmin,
arXiv:1112.5637 [hep-th]

\bibitem {Castellani}L. Castellani, Symmetries in Constrained Hamiltonian
Systems, Ann. Phys. \textbf{143} (1982) 357-371

\bibitem {HTZ}M. Henneaux, C. Teitelboim and J. Zanelli, Nucl. Phys. B 332
(1990) 169

\bibitem {2D}N. Kiriushcheva, S. V. Kuzmin and D.G.C. McKeon, Mod. Phys. Lett.
A 20 (2005) 1895
\end{thebibliography}
\end{document}